\newcommand\les{^{<}_{\sim} }

\newcommand\refind{\hbox to 0.5 truein {\hrulefill .}}                          
\newcommand\beq{\begin{equation}}                                               
\newcommand\eeq{\end{equation}}    

\newcommand\vs{\vskip 0.15 truecm }

                 \documentclass[12pt]{article}
                            \usepackage[english]{babel} 
                            \usepackage{color}
  \usepackage{amsmath}
 \usepackage{graphicx}                                    
  \usepackage{latexsym}          

\begin{document}
  
    \pagestyle{plain}
 
 \centerline{\bf\Large  General Relativity}
\vskip 2truecm
 \centerline{\bf V. M. Canuto$^1$ and I. Goldman$^{1,2}$}
 \vskip 3truecm
 \begin{center}
  {\bf $^1$NASA Goddard Space Flight Center,  Institute for Space Studies 
    2880 Broadway  New York, NY 10025, email: vittorio.m.canuto@nasa.gov}
  \vskip 1truecm
   {\bf   $^2$Department of Exact Sciences, Afeka College, 38 Mivtza Kadesh Tel Aviv  6998812, Israel, email: goldman@afeka.ac.il}
  
  \end{center}
 \vskip 3 truecm 
   
\begin{abstract}
 This is an English translation of the Italian version of an encyclopedia chapter that appeared  in the Italian Encyclopedia of the Physical Sciences, edited by Bruno
Bertotti (1994). Following requests from colleagues  we have decided to make
it available to a more general readership.

We present the motivation for  constructing General Relativity, provide a short discussion of tensor algebra, and follow the set up of Einstein equations. We discuss briefly the initial value problem, the linear approximation and how should non gravitational physics be described in curved spacetime.
\end{abstract}

\section*{1. Introduction: The Need for   a Relativistic Gravity Theory}

The special Theory of Relativity (SR)proposed in 1905 by Einstein, has brilliantly solved many of the severe problems faced by physics at the turn of the 20th century. SR successfully provided a Lorentz  covariant formulation for both particle dynamics and for the the electromagnetic field. Thus a covariant formulation in all inertial frames was achieved. 
\vs
However, SR fell short of answering two central questions:
\vs
(a) How should be physics formulated in non inertial reference frames, i.e. accelerated frames.
\vs
(b) How should gravity be reconciled with SR. Clearly, Newtonian gravity, where the gravitational potential $\phi$ satisfies Poisson's equation $\nabla^2 \phi = - 4 \pi G c^{-2} \rho$, is not Lorentz  covariant(here $G$ is Newton's constant, $\rho$ the matter density and $c$ the speed of light). This fact is in sharp contrast with Maxwell's equations for the electromagnetic field which {\it are} Lorentz covariant.
\paragraph{•}
A theory providing answers to both questions was proposed by Einstein in 1915, the General Theory of Relativity (GR). It took Einstein ten years of hard work (as we learn from  his writings) to bridge the gap and develop GR. Gravity, in GR, is described by means of a curved four dimensional Riemannian spacetime. The  Minkowski flat spacetime of SR is a special case which applies globally only when gravity is absent.

\paragraph*{•}
GR is covariant under general coordinate transformations, and also provides covariant description of physics in general coordinate systems. Thus, GR fully answers question (a) above. With  regard to question (b) above the answer is quiet interesting. While being a relativistic theory , GR is {\it not} contained  within the framework of SR (contrary Maxwell equations). Rather, it implies that SR is incompatible with global  gravity. 
It is of interest to note that all attempts at constructing a relativistic gravity theory within   SR have resulted in either inconsistencies or in contradictions to experimental and observational data.
  
\paragraph*{•}
The predictions  of GR regarding planetary motions in the solar system, compact stellar objects, gravitational radiation, gravitational lensing and cosmology, are in accord with the observations. The most dramatic observational confirmations were made possible in the last decades. Today, GR is regarded as the most elegant and successful physical theory ever proposed.

\section*{2. Geometrical Approach to Gravity}
\subsection*{2a. The Equivalence Principle}

The Equivalence Principle (EP) played a central role in the development of GR. As originally formulated by Einstein in 1908, it states: "{\it all effects of a uniform gravitational field are identical to the effects of a uniform acceleration of the coordinate system}``.

This implies that a uniform gravitational field can be transformed away by an appropriate accelerated reference frame, i.e. one that is {\it freely falling} in the field. In this inertial frame, SR applies and all physical laws retain their SR form.
The EP suggests that the two questions posed in Chapter 1 are interconnected and may refer to two aspects of the same fundamental entity.

In the case of a non uniform field, The EP can be applied {\it locally} in a sufficiently small domain of space and time, so that the variation of the  field can be ignored. Thus, one can introduce a local inertial  Lorentzian  reference frame in which {\it all} physical laws retain their SR form. This formulation of the EP is referred to as "the Strong  EP" (SEP) distinguishing it from other formulations. For more details see articles 7 and 8.

\begin{figure}
 \includegraphics[scale=0.5]{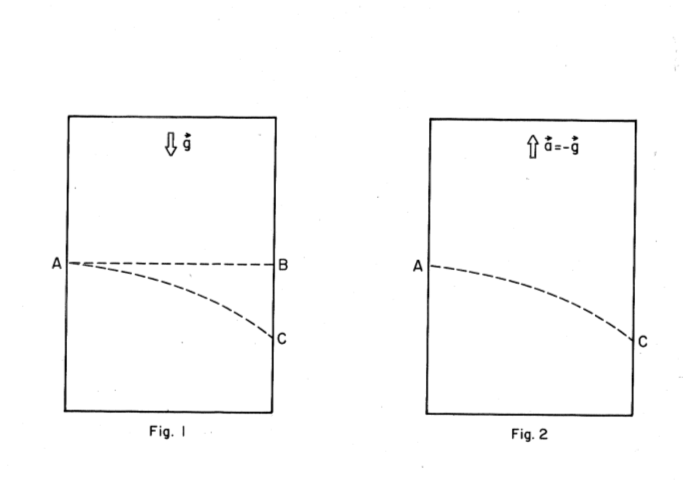}
\end{figure}

The EP has some immediate   implications:
\begin{itemize}
\item[(1)] The ratio of gravitational to inertial mass of a body is a universal dimensionless constant resulting in the universality of free fall.

\item[(2)]
Light rays passing nearby a massive body follow bent trajectories. 

\item[(3)]  

A photon traveling up (down) in a gravitational potential   undergoes a gravitational red (blue) shift.

Let us discuss these points:

\item[(1)] The gravitational mass of a body $m_g$ is defined through the gravitational force equation $\vec F_g = - m_g \nabla\phi$ is playing the role of "gravitational charge".The inertial mass of the body $m_i$ is the one appearing in Newton's second law $\vec F = m_i \vec a$.
According to the EP, a given gravitational field can locally be replaced by an accelerated reference frame. Thus, the ratio $m_g/m_i$ must be the {\it same} for all objects. This universality has been tested by Etvoes, Pekar and Fekete (1922) who found that $\frac{\delta (m_g/m_i)}{(m_g/m_i)} \leq  5\times 10^{-9}$ and later by Roll, Krotkov and Dicke (1964) ($\leq 10^{-11}$) and Braginsky and Panov (1972) ($\leq 10^ { -12}$).

\item[(2)]
Consider a box located in a uniform gravitational field (Fig. 1).
A photon is emitted in the horizontal direction from A. Will it move along  line AB or rather along a bent down curve AC,  as a massive body will? To answer this, consider an identical box    which  is gravity free but is accelerated upward  with an acceleration  $ \vec a= -\vec g$ ( Fig. 2). According to the EP, physics should be the same  in both boxes. Let us emit a photon in the horizontal direction in the  box of Fig.2. In this case the trajectory will be bent down parabola AC, since the box is accelerating upward in a free force  region. The EP then implies that in Fig. 1, the photon will {\it also} follow a bent down trajectory AC.
Bending of starlight grazing the sun's surface was first observed in a 1919 eclipse. Substantially better confirmations were obtained in recent decades by measurements of deflection of radio waves emitted by QSOs in the solar gravitational field (Counselman et. al. 1974). Gravitational lensing, based on the trajectory bending of photons,  has become in recent years  an important astronomical observational tool.

\item[(3)]  

Consider once more boxes (a) and (b). Let a photon be emitted from the bottom  and detected at the top. In box (b) one can compute the Doppler redshift by noting that, at the moment the photon reaches the top, the bottom has a relative velocity $ v = |\vec a| h / c$. Thus the redshift is $z \equiv \lambda_{top}/\lambda_{bottom}= 1 + v/c$. According to  the EP this is also the redshift in box (a). Now, since $|\vec a|= |\vec g|$, we have $z= 1+ \frac{\Delta \phi}{c^2}$ whee $\Delta \phi$ is the gravitational potential difference between the top and the bottom. This gravitational redshift formula was verified to a precision of 1 part in 100 by Pound and Rebka (1960) and by Pound and snider (1965) who used the Mossbauer effect to detect $\gamma$ rays from $Fe^{57}$ caused by the height difference  of 22.5 meters in the terrestrial gravitational field. A verification with a precision of 7 parts in $10^5$ was obtained by Vessot, Levine, et al. (1980) by monitoring the frequencies of two hydrogen masers clocks:  one on board of a rocket launched to an altitude of $10^4$ km and the other on Earth surface.
 \end{itemize} 
\subsection*{2b. The Need for a Metric Gravity Theory as a Result of combining SR and EP}

As demonstrated in the previous subsection, combining the SR and EP can yield interesting results regarding gravity. One can use the same approach in order to obtain an insight into the general approach towards developing a relativistic gravity theory. In order to do so, let us summarize  some of the main points of SR.

\paragraph{(i)}
 
Existence of a Lorentz-invariant  squared line element $ ds^2$,  in a four dimensional Minkowski spacetime.

 $$ds^2= \eta_{\mu\nu}dx^{\mu} dx^{\nu}\ \ ; \ \ \ \ \ \ \ \eta_{\mu\nu} = diag(1, -1, -1, -1) \eqno(2.1)$$
 
   The line elements measures proper time intervals and proper physical distances  between two neighboring events.

\paragraph{(ii)}

Massive particles move along timelike($ ds^2>0$) trajectories. Free particles move along timelike  straight lines

$$\frac{d^2x^{\mu }}{ds^2} =0\ \ ; \ \ \ \ ds^2 > 0    \eqno(2.2)        $$

\paragraph{(iii)} Photons (and other zero mass particles) move along null straight lines.

$$\frac{d^2x^{\mu}}{dp^2} =0\ \ ; \ \ \ \ ds^2 = 0    \eqno(2.3)        $$
where  $p  \neq  s$ is the line  parameter.
\paragraph{(iv)}

All physical quantities are represented by tensors in the Minkowski spacetime.

Consider now a region of spacetime permeated by a gravitational field. Let us use a coordinate system
$\{x^{\mu}\}$ to mark four points in the region. Apply the EP to transform to a local Lorentz coordinate system $\{\bar x^{\mu}\}$. In this frame

$$    \bar{ d s^2}  = \eta_{\mu\nu} d \bar x^{\mu} d \bar x^{\nu}  \eqno(2.4) $$

Let us express

$$\bar x^{\mu} = F^{\mu}( x^{\mu})  $$

so that $d\bar x^{\mu} = F^{\mu}_{,\alpha} dx^{\alpha}$. Assume further that the  line element squared, being a scalar, is invariant under general coordinate transformations,   and not only under inertial coordinate  transformations.
\vfill\eject
 This yields

$$\bar{ d s^2} = ds^2 =\eta_{\mu\nu}  F^{\mu}_{,\alpha}   F^{\nu}_{,\beta}dx^{\alpha}dx^{\beta} = g_{\alpha\beta} dx^{\alpha}dx^{\beta}\eqno(2.5)$$

where

$$  g_{\alpha\beta} = F^{\mu}_{,\alpha}   F^{\nu}_{,\beta} \eqno(2.6)$$

Thus when expressed in the original coordinates $\{x^{\mu}\}$, $ ds^2$ has the form of a squared line element in a curved four dimensional Riemannian space.
This suggests that a gravitational field can be represented by a geometry of a curved four dimensional spacetime. The emerging geometric approach to gravity   can be summarized as follows:

\paragraph{\hskip  1 truecm (a)} Physical spacetime is a four dimensional Riemannian space. In gravity free regions, the space is a flat Minkowski spacetime and SR holds.

\paragraph{\hskip  1 truecm (b)} At any four point in spacetime it is possible to
introduce  a local inertial Lorentzian coordinate system in which SR  holds, and local physics retains it SR form.
\paragraph{\hskip  1 truecm (c)} Laws of physics should be formulated in a general covariant manner. This implies the physical quantities should be represented by tensors in a curved Riemannian space. 
 \paragraph{\hskip  1 truecm (d)}The line element squared defines proper spacetime intervals between neighboring events. As such , it should be invariant under general coordinate transformations.
 
 Using the above principles and the EP, one can conclude the following results:

  \paragraph{\hskip  1 truecm (e)} Massive particles move along timelike  ($ ds^2>0$) trajectories. Free particles (not subject to any non gravitational interaction) move along timelike geodesics.
  \paragraph{\hskip  1 truecm (f)} 
 Photons (and other zero mass particles) move along null geodesics.
 
 see 2d. for a derivation of (e) and (f) above.

\subsection*{2c. Coordinates, Metric and Spacetime Measurements} 
  
Coordinates are any set of four dimensional markers, attached continuously to spacetime points. As  such,  they need not to have any measurable physical meaning.

As stated in the previous section, measured physical spacetime intervals are given by the line element. Once a coordinate system is chosen, one can determine the components of the metric tensor $g_{\mu\nu}$ by carrying out spacetime measurements. Examples follow.

\subsection*{Measuring   $\bf{g_{00}}$ }

In order to measure $ g_{00}$ at a given spacetime point, consider a clock at  rest in the spatial coordinates $\{x^i\}$. Compare two events on the clock’s world line: $( x^0, x^i), \ ( x^0 + dx^0, x^i )$. In this case $ds^2 = g_{00}\   dx^{0{^2}} $. Let us transform to a local coordinate system 
$\{\bar x^{\mu}\}$ in which the clock is momentary at rest. Thus, in this frame the same invariant
$$ ds^2 = d\bar x^{0^{2}} = d\tau^2 \eqno(2.7)$$
where $d\tau$ is the proper time interval as measured by the clock (in the inertial rest frame). Therefore

$$ ds^2 = g_{00}dx^{0{^2}} = d\tau^2 \eqno(2.8)$$
and so
$$g_{00} = \frac{d\tau^2}{dx^{0{^2}}} \eqno(2.9)$$
is expressed as the square of the ratio between the proper time interval and the corresponding coordinate time interval.

\subsection*{Measuring $\bf{ g_{0i}, \ g_{ij}}$} 

Let two photons be emitted from $(x^0, x^1, x^2, x^3)$ along opposite directions of $x^1$  so that they are detected at $(x^0 +dx^0, x^1 + dx^1, x^2, x^3)$ and
 $(x^0 + dx^{0^*}, x^1 - dx^1, x^2, x^3)$. photons move along null geodesics so for the  pair of events
 
 $$ds^2 = g_{00} dx^{0^2}   + 2 g_{01} dx^1dx^0 + g_{11}dx^{1^2} ; \ \ \ \ ds^{*^2} = g_{00} dx^{*^{0^2}}  - 2 g_{01} dx^1dx^*{^0} - g_{11}dx^{1^2} \eqno(2.10)$$  
 resulting in
 
 $$g_{11} = - g_{00}\frac{dx^0 dx^{*^0}}{dx^{1^2}}; \ \ \ \ \ g_{01} = - g_{00}\frac{dx^0 -dx^{*^0}}{2 dx^1 }  \eqno(2.11) $$
 
 Therefore, given the above measured $g_{00}$ ,  and the coordinate intervals $ dx^1, dx^0, dx^*{^0}$, one can determine $g_{11}$ and $g_{01}$. The procedure for measuring $g_{0i}$ and $g_{ii} ( i= 2, 3)$ is the same. By similar procedures, one can measure $g_{ij} (i\neq j)$. 
 
 \subsection*{2d.  Geodesic Equations of Motion}
 
 What trajectories follow material particles,   which are free from all other interactions, in a gravitational field? To answer this question, consider a region of spacetime in which there exists a gravitational field. Let us use the coordinates $\{x^{\alpha}\}$. In a local inertial coordinate system $\{\bar x^{\alpha}\}$, SR holds so that free particles move along straight timelike lines
 
$$\frac{d\bar x^{\mu} }{ds^2} = 0\ \ ; \ \ \ \ ds^2>0 $$ 
 
 Employing the EP and expressing $\bar x^{\mu}$ by $x^{\alpha}$ as in section 2b, one finds, after some algebra 
 
 $$g_{\mu\alpha} \frac{d^2x^{\alpha}}{ds^2} + \left( g_{\mu\alpha,\beta} - \frac{1}{2} g_{\alpha\beta,\mu}\right) \frac{dx^{\alpha}}{ds}\frac{dx^{\beta}}{ds} \ \ ; \ \ \ \ ds^2>0     \eqno(2.12)$$ 
 
 where $g_{\mu\nu}$ is given by Eq.(2.6).
 
 Equation (2.12)describes a timelike  geodesic in the curved geometry defined by 
 $g_{\mu\nu}$. It can be easily shown that a geodesic is also the curve that extremizes the total arc length, $\int ds$ , between two four points in spacetime.
A more formal definition of a geodesic will be introduced in chapter 3.

 As for a non free particle, the  equations of motion will not be those of a geodesic, however the line element is timelike  $( ds^2 > 0)$. This follows immediately from the EP and SR.
 
 To obtain the trajectory of a photon in a gravitational field, recall that  in the local inertial coordinates the trajectory is given by

$$\frac{d\bar x^{\mu} }{dp^2} = 0\ \ ; \ \ \ \ ds^2 =0 $$ 
 
 where $p \ne s$ is the curve parameter. Following the same steps as in the case of massive particle , the result is
 $$g_{\mu\alpha} \frac{d^2{x^\alpha}}{dp^2} + \left( g_{\mu\alpha,\beta} - \frac{1}{2} g_{\alpha\beta,\mu}\right) \frac{dx^{\alpha}}{dp}\frac{dx^{\beta}}{dp} \ \ ; \ \ \ \ ds^2=0     \eqno(2.13)$$ 
This is an equation of a null geodesic in the curved geometry defined by 
 $g_{\mu\nu}$. 
 
 \paragraph*{Correspondence with Newtonian Gravity}\hfill
  
 In order to establish a correspondence between Newtonian gravity and the geometric description of relativistic gravity, it is useful to examine the geodesic equations of motion for slow particles ($v<<c$) in a weak, static gravitational field.

A static field corresponds to a metric of the form:

$$ds^2 = g_{00} dx^{0^2} + g_{ij} dx^i dx^j\ \ ; \ \ \ \ g_{\mu\nu ,0}=0 \eqno( 2.14)$$ 
 
 in the case of a weak field 
 $$g_{\mu\nu} = \eta_{\mu\nu} + h_{\mu\nu}\ \ ; \ \ \ |h_{\mu\nu}|<<1 \eqno(2.15)$$
 Using Eqs.( 2.14), (2.15), the geodesic equations(2.12) become
 
 $$\frac{d^2x^0}{ds^2} = 0 + O\left( v/c\  h_{\mu\nu}\right) $$
 implying
 
 $$\frac{d x^0}{ds } = 1 + O\left( (v/c)^2\right) +O\left(  h_{\mu\nu}\right)\eqno(2.16)$$
 
 $$\frac{d^2x^i}{ds^2} =- \frac{1}{2}h_{00, i}  + O\left( v/c\  h_{\mu\nu}\right) $$
hence

 $$\frac{d^2x^i}{{dx^0}^2} =- \frac{1}{2}h_{00, i}  + O\left( v/c\  h_{\mu\nu}\right) \eqno(2.18)$$

 Eq.(2.18) should be compared with the Newtonian equations of motion
 $\frac{d^2x^i}{{dt}^2} =-\phi_{, i}$, where $\phi$ is the gravitational potential. 
 Since $x= ct$, we may identify
 $$h_{00} = \frac{2 \phi}{c^2}\ , \eqno(2.19)$$
 namely the deviation of $g_{00}$ from unity is proportional to the gravitational potential $\phi$.

 \subsection*{2e.  Gravitational Redshift}
 
 Consider a photon emitted at $\{x_e^{\mu}\}$ and detected at $\{x_d^{\mu}\}$.
 The frequencies of the photon measured by the emitter and by the detector are in general different. The difference depends on the world-lines of the emitter and the detector and on the gravitational field in the region under consideration. Let us focus here on the effects of the gravitational field. Consider therefore,  a static field with the line element squared given by Eq.(2.14) and emitter and detector at rest in the coordinate system. Since the photon  trajectory is a null geodesic it follows   that
 
 $$x_d^0 - x_e^0 = F( x_e^i, x_d^i) \eqno(2.20)$$ 
 
 Consider a pulse of N photons with  duration of N periods which is being emitted and detected. From Eq.(2.20) it follows that ($dt$ denoting $dx^0$)
 
 $$   dt_d = dt_e\ ,$$
 where the time intervals are coordinate time intervals corresponding to N periods, at the detector and emitter respectively.
 
 Now, for an emitter and detector at rest at the moments of emission and detection respectively, the periods are given by
 
 $$P_e= \frac{1}{N}d\tau_e= \frac{1}{N}{ g_{00}}_e^{1/2}dt_e\ ; \ \ \  P_d=\frac{1}{N}d\tau_d= \frac{1}{N} {g_{00}}_d^{1/2}dt_d \eqno(2.21)$$
 hence the redshift, $Z$ is
 
 $$z \equiv \frac{\lambda_d}{\lambda_e} = \sqrt{\frac{{g_{00}}_d}{ { g_{00}}_e  }} \eqno(2.22)$$
 
In the case of a weak field eqs. (2.19) and (2.22) imply that

 $$z = 1+ \frac{\phi_d - \phi_e}{c^2} \eqno(2.23)$$
 which coincides with the expression for $z$ obtained in section 2a, using the EP.
 
 \section*{3. Tensor Algebra and Analysis- a Short Review.}
 As already stated  in section 2b, general covariance implies that physicality quantities be represented by tensors in a four dimensional Riemannian space. A brief summary of tensor algebra and analysis follows.
 
 \subsection*{3a. Transformation laws }
 
 For a physical quantity defined as function of spacetime four points to be a tensor, it must undergo a specific transformation under the coordinates transformation $ x^{\mu} \to \bar x^{\mu}(x^{\alpha})$.
 
 \paragraph*{Scalar}
 
 A one component quantity whose value is unchanged under coordinates transformation
 
 $$\bar\phi(\bar x^{\mu}) = \phi(x^{\mu})\eqno(3.1) $$
  \paragraph*{Contravariant Vector}
 
 four component quantity transforming in the same way as $dx^{|mu}$ 
 
 $$\bar x^{\mu}= \frac{\partial\bar x^{\mu}}{\partial x^{\alpha} } A^{\alpha}   \eqno(3.2)$$
 
 \paragraph*{Covariant Vector}
 
 four quantity transforming in the same way as $\frac{\partial}{\partial x^{\mu}  }$
 
 $$\bar x_{\mu}= \frac{\partial   x^{\alpha}}{\partial\bar x^{\mu}} A_{\alpha}  \eqno(3.3)$$
 
 \paragraph*{Tensor}
 
Multi indexed quantity; number of components equals four times the number of induces. Each contravariant (upper) index and each covariant (lower) index 
 transform as does the corresponding vector type.
 
 $$ \bar T_{\alpha \cdots \beta}^{\mu \dots\nu}= T_{\gamma \cdots \delta}^{\theta \cdots\rho} \cdots\frac{\partial \bar x^{\mu}}{\partial x^{\theta}}  \cdots  \frac{ \partial \bar x^{\nu}}{\partial x^{\rho}} \cdots \frac{\partial x^{\gamma}}{\partial\bar x^{\alpha}} \cdots  \frac{\partial x^{\delta}}{\partial\bar x^{\beta}}  \eqno(3.4)$$
  
 \subsection*{3. Algebra }
 
  \paragraph*{Raising and Lowering Indices}
 
Given a contravariant vector $A^{\mu}$, one may define an associated covariant vector  $A_{\mu}$ 
 by
 $$A_{\mu} \equiv g_{\mu\nu}A^{\nu}\eqno(3.5)$$
 
Similarly, given a covariant vector $A_{\mu} $, one may define an associated contravariant  vector  
 $A^{\mu}$ 
by 
$$ A^{\mu} =   g^{\mu \nu} A_{\nu} \eqno(3.6)$$ 
where $g^{\mu \nu}$ is the inverse of  $g_{\mu \nu}$
 
 $$g_{\mu\nu} g^{\nu\alpha} = \delta_{\mu}^{\alpha} \eqno(3.7)$$
 These procedures  are readily generalized to any contravariant and covariant index of a general tensor.
 
   \paragraph*{Scalar Product}
 
The scalar (inner) product of two vectors is the sum of the products of the contravariant component by the corresponding covariant component.

$$\phi \equiv A^{\mu} B_{\mu} = g_{\mu\nu} A^{\mu}B^{\nu} \eqno(3.8)$$

The result is a scalar as can be easily verified by checking its behavior under a coordinate transformation (using the known transformation rules of $A^{\mu} $ and $B_{\mu}$). One can generalize the scalar product operation to   apply to two tensors. The resulting tensor is of a degree equaling the sum of the degrees of the multiplying tensors minus 2. E.g. $R^{\mu\nu\alpha} A_{\alpha}$ i s a second order contravariant tensor.
 
    \paragraph*{Norm of a Vector}
    
    This the scalar product of the vector with itself.
   $$ A^{\mu} A_{\mu} = g_{\mu\nu} A^{\mu}A^{\nu} \eqno(3.9)$$
   
     \paragraph*{Contraction}
     
     Contraction is the inner summation over a repeated index of a mixed tensor, resulting in a tensor of a degree lower by 2 than that of the original tensor, e.g., $B_{\nu}^{\mu\nu}$ is a contravariant vector. 
   
  \subsection*{3c. Analysis}

    \paragraph*{Tensor Field}
    
    Is an assignment of a tensor at each   four point of a given domain of spacetime. The components of the tensor field are functions of position in spacetime.

    \paragraph*{Parallel transport of Contravariant Vectors}
    
    Consider  a vector $A^{\mu}$ at a four-point $x_{0}^{\alpha}$. Let us pass through that point a curve, $x^{\alpha}(p)$, with $p$ the curve parameter. We want to transport the vector along the curve so as to form the closest possible generalization to a constant vector field along a curve in flat spacetime.The transport law should be linear in $A^{\alpha}$ since one   would like that the transport of a sum of two vectors be equal to the sum of the two transported vectors. We demand also that the transported vector is parallel to the curve.
Let's write therefore

$$\frac{dA^{\mu}}{dp} = -  \Gamma_{\alpha\beta}^{\mu}  \frac{dx^{\beta}}{dp}A^{\alpha} \eqno(3.10)$$    
and demand that
   
\paragraph{(i)}

 $A^{\mu} + dA^{\mu}$ is a vector at the point $x^{\alpha}(p+dp) = x^{\alpha} + dx^{\alpha}$.

\paragraph{(ii)}
That for any four point one can find a local Lorentzian coordinate system in which $\Gamma_{\alpha\beta}^{\mu}$
vanish at this point. Thus, in the point's local neighborhood, Eq.(3.10) results in a constant vector field.
 
\paragraph{(iii)}
That the scalar product of two transported vectors, and in particular the norm of a transported vector, remains constant along the path $x_{0}^{\alpha}$.

These requirements determine uniquely $\Gamma_{\mu\nu}^{\alpha}$

$$\Gamma_{\mu\nu}^{\alpha}  = \frac{1}{2} g^{\alpha\beta}\left( g_{\mu\beta, \nu} +  g_{\nu\beta, \mu} - g_{\mu\nu, \beta}\right) \ , \eqno(3.11)$$

as well as it's transformation law

$$ \bar\Gamma_{\mu\nu}^{\alpha}  =  \Gamma_{\beta\delta}^{\sigma}\frac{\partial \bar x^{\alpha}}  {\partial x^{\sigma}} \frac{\partial  x^{\beta}}{\partial\bar x^{\mu} }\frac{\partial  x^{\delta}}{\partial\bar x^{\nu} } - \frac{\partial^2\bar x^{\alpha}}{  \partial x^{\beta}\partial x^{\delta} }\frac{\partial x^{\beta}}{\partial\bar x^{\mu}} \frac{\partial  x^{\delta}}{\partial\bar x^{\nu} } \eqno(3.12)$$

Thus  $\Gamma_{\mu\nu}^{\alpha}$, known as the Christoffel symbol , is {\it not} a tensor.
We note that the transported value of the vector $ A^{\mu}$ at a given point {\it depends} on the curve joining the point with the original point  where the original vector was defined. Thus in a general spacetime, one cannot generate a vector field by parallel transport. We discuss this   in some detail in chapter 5.

\paragraph{Parallel Transport of a Covariant Vector}

By using the fact that the scalar product of parallel transported vectors is constant along the curve

$$ \frac{d}{dp}( A_{\mu} B^{\mu})= \frac{d A_{\mu}}{dp }B^{\mu} + A_{\mu}  \frac{dB^{\mu}}{dp }   =0 $$ 
and applying Eq.(3.10), it follows that ( because the two original vectors are general) 

$$ \frac{dA_{\mu}}{dp} = \Gamma_{\mu\beta}^{\alpha} A_{\alpha} \frac{dx^{\beta}}{dp} \eqno( 3.13) $$
\paragraph{Covariant Derivative}
Given a vector field $A^{\mu}(x^{\alpha}$), it is easy to check that $A^{\mu}_ ,\nu$ does not transform as a tensor. One can form a derivative that is a tensor by taking, at $x^{\alpha} + dx^{\alpha}$, the difference between $A^{\mu}( x^{\alpha} + dx^{\alpha})$ and the parallel transported value of $A^{\mu}(x^{\alpha})$ at  $x^{\alpha}$. The difference is    

$$ {A^{\mu}}_{,\nu} dx^{\nu}   + \Gamma_{\alpha\nu}^{\mu} A^{\alpha}dx^{\nu}      $$
 The resulting covariant derivative 

$${A^{\mu}}_{;\nu}  =  {A^{\mu}}_{,\nu}   + \Gamma_{\alpha\nu}^{\mu} A^{\alpha} \eqno ( 3.14) $$

is a tensor.   The covariant derivative reduces to the ordinary partial derivative in a local inertial  coordinate system, since in the latter $\Gamma_{\alpha\nu}^{\mu} =0$.
In the case of a covariant vector the covariant derivative is given by

$${A_{\mu}}_{;\nu}  =  {A_{\mu}}_{,\nu}  - \Gamma_{\mu\nu}^{\alpha} A_{\alpha} \eqno (3.15) $$

The covariant derivative of a general tensor is given by,

 $$\left(T_{\alpha \cdots \beta}^{\mu \dots\nu}\right)_{;\lambda} = \left(T_{\alpha \cdots \beta}^{\mu \cdots\nu}\right)_{,\lambda} + \Gamma_{\theta\lambda}^{\mu} T_{\alpha \cdots \beta}^{\theta  \cdots \nu}  \cdots   +   \Gamma_{\theta\lambda}^{\nu} T_{\alpha \cdots \beta}^{\mu \cdots\theta}  \cdots -   \Gamma_{\alpha \lambda}^{\theta} T_{\theta \cdots \beta}^{\mu \cdots\nu}  \cdots -   \Gamma_{\beta \lambda}^{\theta} T_{\alpha \cdots \theta}^{\mu \cdots\nu}  \eqno(3.16)   $$

The covariant derivative of a product is expanded in the same way as an ordinary partial derivative of a product, e.g.

$$ \left( T^{\mu\nu} B_{\alpha}\right)_{; \lambda} = T^{\mu\nu}_{;\lambda} B_{\alpha} + T^{\mu\nu}  B_{\alpha;\lambda}   \eqno (3.17)$$

The covariant derivative of the metric  tensor vanishes identically

$$ g_{\mu\nu;\alpha} =0 \eqno(3.18)$$
 
\paragraph{Divergence of a Vector}

 $$ A^{\alpha}_{;\alpha} = A^{\alpha}_{,\alpha} + \Gamma _{\alpha\beta}^{\alpha} A^{\beta} = \frac{\left( \sqrt{-g}A^{\alpha}\right)_{,\alpha }}{\sqrt{-g}}\eqno (3.19)$$

 where $g \equiv det(g_{\mu\nu})$.
 
\paragraph{Curl of a Vector}
 
 Since $\Gamma_{\mu\nu}^{\alpha} = \Gamma_{\nu\mu}^{\alpha}$ we find
 
 $$A_{\mu;\nu} -A_{\nu;\mu} = A_{\mu,\nu} -A_{\nu,\mu} \eqno(3.20)$$
 
 \paragraph{Geodesics}
 
 A geodesic is defined as a curve  $ x^{\mu}(p)$ along which the tangent to the curve $\frac{d x^{\mu}}{dp}$ is being parallel transported. This definition is the generalization of the concept of a straight line in a flat space. Thus,
 
 $$ \frac{d}{dp}\left(\frac{dx^{\mu}}{dp} \right) = -\Gamma_{\alpha\beta}^{\mu} \frac{dx^{\alpha}}{dp}\frac{dx^{\beta}}{dp} \eqno(3.21) $$
 Comparison with Eq.(2.12) shows (after contracting Eq.(2.12) with $ g^{\mu\nu}$)
 that the two are indeed identical.
 
 Since the norm of a parallel transported vector is constant along the curve, one has in the case of a geodesic 
 
 $$\left(\frac{ds }{dp}\right)^2 = g_{\mu\nu} \frac{dx^{\mu}}{dp}\frac{dx^{\nu}}{dp}= const$$.
 
  Hence, a geodesic retains its timelike , spacelike, or null character along its full extent.
 From Eq.(3.21) it follows that any other parameter $q$ satisfying it, must be related to $p$ by $\frac{dp}{dq}= const.$ Consider a non null geodesic.  it satisfies 
 
  $$\left(\frac{ds }{dp}\right)^2 = g_{\mu\nu} \frac{dx^{\mu}}{dp}\frac{dx^{\nu}}{dp}= const$$. 
  
  Hence $\frac{dp}{ds}= const.$, and $s$ can serve as the geodesic parameter. For null geodesic $ds=0$ and one must have $p\neq s$. 
 
 \subsection*{3d. Tensor densities}
 
 Under a coordinate transformation $x^{\mu}  \rightarrow \bar x^{\mu}( x^{\alpha})$, the determinant of the metric tensor $ g\equiv det(g_{\mu\nu})$ transforms according to
 $$\bar g = g J^{-2} \eqno(3.22)$$
 where 
 $$ J \equiv det\left(\frac{\partial \bar x^{\alpha}}{\partial   x^{\beta} }\right)  \eqno(3.23) $$
 
is the Jacobian of the transformation.

The four  volume element $d^4x$ transforms like

$$d^4\bar x  = J d^4 x \eqno(3.24) $$ 
 
 Thus
 
 $$\sqrt{-\bar g} d^4 \bar x = \sqrt{-  g} d^4 x \eqno(3.25)$$
 is a scalar.

 Given a tensor $T^{\mu\nu}$ one defines the associated tensor density $\tau^{\mu\nu}$ by
 
 $$\tau^{\mu\nu}\equiv \sqrt{-g}T^{\mu\nu} \eqno(3.26)  $$
 
 From Eq.( 3.25) it follows that $\tau^{\mu\nu}d^4 x$ is a tensor.
 Integrals of tensor densities , however have no definite transformation laws and hence are {\it not } tensors. To see it note that if
 
 $$A^{\mu\nu} \equiv \int \tau^{\mu\nu} d^4x = \int \sqrt{-g}T^{\mu\nu}d^4x \eqno(3.27)$$
 then 
 
 $$\bar A^{\mu\nu}  = \int \sqrt{-\bar g}\bar T^{\mu\nu}d^4\bar x = \int \sqrt{-g}T^{\alpha\beta} \frac{\partial \bar x^{\mu}}{\partial x^{\alpha}}  \frac{\partial \bar x^{\nu}}{\partial x^{\beta}}  d^4x
  \eqno(3.28)$$
 
 Only for linear transformations where $\frac{\partial\bar  x^{\mu}}{\partial x^{\alpha}}$ are constants does $A^{\mu\nu}$ behave like a tensor.
 
Tensor densities and their integrals will be encountered in Chapters 6, 7 and 8.

\section*{4. Local Inertial Frames} 
 
 One of the principles of the geometric approach to gravity (section 2b) is the existence, at any spacetime point of a local inertial coordinate system in which locally $g_{\mu\nu} = \eta_{\mu\nu}$ and SR applies. In this section we explicitly construct such a system and quantify its properties.
 
Given a coordinate system $\{x^{\mu}\}$ let us construct a local inertial coordinate system at the vicinity of the four point $x_p^{\mu}$ by defining the coordinates $\{ \bar x^{\alpha}\} $:

 $$ \bar x^{\alpha}= x^{\alpha} - x^{\alpha}_p +\frac{1}{2}\left(\Gamma_{\mu \nu}^{\alpha}\right)_p\ \left(x^{\mu} - x^{\mu}_p\right) \left(x^{\nu} - x^{\nu}_p\right)  \eqno(4.1) $$

The point  $x_p^{\mu}$ is the origin of the   $\{ \bar x^{\alpha}\} $ 
 coordinate system since $ \bar x^{\alpha} \left( x_p^{\mu} \right) = 0 $. 
 Furthermore, from Eq.(4.1) one finds

 $$\left(\frac{\partial\bar x^{\alpha}}{\partial x^{\mu}}\right)_p =  \delta_{\mu}^{\alpha}\ \ \ , \ \ \ \  \left(\frac{\partial^2\bar x^{\alpha}}{  \partial x^{\mu}\partial x^{\nu}}\right)_p  = \left(\Gamma_{\mu \nu}^{\alpha}\right)_p  \eqno(4.2) $$ 
 
 which , upon applying the transformation law for $g^{\mu\nu}$, yields
 
 $$\left(\bar g^{\mu\nu}\right)_p =  \left(  g^{\mu\nu}\right)_p \ \ {\rm therefore, } \left(\bar g_{\mu\nu}\right)_p =  \left(  g_{\mu\nu}\right)_p  \eqno (4.3)$$
 
By using the transformation law of the Christoffel symbols, Eq.(3.12), and Eq.(4.2) we find that

$$\left(\bar\Gamma_{\alpha\beta}^{\mu}\right)_p =0  \eqno(4.4) $$

implying ( by use of Eq.(3.11))
 
 $$(\bar g_{\mu\nu, \alpha})_p =0  \eqno(4.5) $$

At this stage one may apply, further, a linear transformation 

$$ \widetilde x^{\alpha} = \Lambda_{\mu\nu}^{\alpha} \bar x^{\mu};\ \ \ \ \Lambda_{\mu\nu}= constants \eqno(4.6) $$ 

to diagonalize and normalize $(\bar g_{\mu\nu})_p$, resulting in 

$$ (\widetilde g_{\mu\nu})_p = \eta_{\mu\nu} \eqno(4.7)$$

while keeping

 $$(\widetilde g_{\mu\nu, \alpha})_p =0 \ \ \ ; \ \ \ \ \left(\widetilde  \Gamma_{\alpha\beta}^{\mu}\right)_p =0  \eqno(4.8) $$
 
Thus in the neighborhood of this system origin  $(\widetilde x^{\alpha})_p = 0\ , \ \ \widetilde g_{\mu\nu}= \eta_{\mu\nu}$ up to the $2_{\rm nd}$  order in $\widetilde x^{\alpha}$.

In the coordinate system  $\{\widetilde x^{\alpha}\}$ one has in a sufficiently small neighborhood of  $(\widetilde x^{\alpha})_p = 0$ :
 
\paragraph{\hskip 1 truecm (i)}  Covariant derivatives reduce to ordinary partial derivatives.

 \paragraph{\hskip 1 truecm (ii)} The equations of geodesics are straight lines
$$\frac{d^2 \widetilde x^{\mu}}{dp^2} = 0 \eqno(4.9) $$ 
 
 \paragraph{\hskip 1 truecm (iii)}
 
$$\widetilde g_{\mu\nu} = \eta_{\mu\nu} + O( \widetilde x^{{\alpha}^2}) \eqno(4.10)   $$ 
 
Thus $\{\widetilde x^{\alpha}\}$ has indeed the characteristics of the local inertial frame. The coordinate system axes themselves satisfy Eq.(4.9) and hence are geodesics. This suggests a practical way of constructing a local inertial frame at any given four point. Pick  four mutually orthogonal geodesics originating from the point (one   timelike  and three spacelike). The coordinates of a point $\widetilde x^{\alpha}$ are defined to be the proper lengths along the axes-geodesics from the origin up to the normal projections of $\widetilde x^{\alpha}$ on the axes. Thus, the spatial origin of the system ($\widetilde x^i = 0$) moves along the free falling timelike  geodesic.

Is it possible to find an inertial coordinate system such that also the second derivatives of the metric tensor vanish at the coordinates origin? The answer is negative: the reason being that the Riemann curvature tensor (see Chapter 5) cannot be made to vanish by use of coordinate transformations. Thus  while one may, via the local inertial frame, eliminate {\it locally} the effects of gravitational field, one cannot do so  {\it globally} in a finite domain of spacetime.
 
\section*{5. Curvature} 

Given a general metric tensor, $g_{\mu\nu}$, one would like to know whether it represents a genuine curved spacetime or just appears to be so because of a particular choice of a the coordinate system.  In physical terms, one would like to know whether a non Lorentz metric represents a gravitational field or is a result of "accelerated" coordinates to describe an otherwise gravity free flat Minkowski spacetime. This question can be answered by examining the Riemann curvature tensor. As will be seen, the vanishing of the latter (which is a coordinate invariant statement) is a necessary and sufficient condition for spacetime flatness. In order to introduce the curvature tensor, let's consider the commutator of the second  covariant derivatives of a vector $A^{\mu}$. Upon using the definition of the covariant derivative, we find 

$$ A^{\mu}_{;\alpha;\beta} -  A^{\mu}_{; \beta ; \alpha} = R^{\mu}_{\rho\alpha\beta}A^{\rho} \eqno(5.1) $$

where

$$  R^{\mu}_{\rho\alpha\beta} \equiv \left(\Gamma^{\mu}_{ \alpha\rho}\right)_{,\beta} -  \left(\Gamma^{\mu}_{ \beta\rho}\right)_{,\alpha} +  \Gamma^{\tau}_{\alpha\rho}   \Gamma^{\mu}_{\tau\beta } -\Gamma^{\tau}_{\beta\rho}   \Gamma^{\mu}_{\tau\alpha} \eqno(5.2) $$
is the Riemann curvature tensor.

\subsection*{Theorem}
 
$R^{\mu}_{\rho\alpha\beta} = 0$ is a necessary and sufficient condition for spacetime flatness.
 \subsection*{Necessary Condition}

Given a flat spacetime, one can introduce a coordinate system in which 
$g_{\mu\nu} =\eta_{\mu\nu}$ globally. Thus $ \left(\Gamma^{\mu}_{ \nu\alpha }\right)_{,\beta} = 0 \ {\rm and}\ \Gamma^{\mu}_{ \nu\alpha} =0$ and from (5.2) it follows that $R^{\mu}_{\rho\alpha\beta} = 0$.  

\subsection*{Sufficient Condition}

To prove that $R^{\mu}_{\rho\alpha\beta} = 0$ is a sufficient condition to ensure spacetime flatness, let us first consider the conditions under which one can generate by parallel transport a true vector field. The latter should be only a function of position in spacetime, independently of the path of transport. For this to occur we must have

$$ A^{\mu}_{,\beta, \lambda} = A^{\mu}_{,\lambda, \beta} \eqno(5.3) $$ 

where $A^{\mu} $ is generated by parallel transport, namely

$$ A^{\mu}_{,\beta} + \Gamma_{\alpha\beta}^{\mu} A ^{\alpha} = 0 \eqno(5.4)$$
Combining eqs.(5.3) and (5.4) one finds, after some algebra, that

$$ R^{\mu}_{\rho\alpha\beta}A^{\rho} = 0   \eqno(5.5) $$

Thus the vanishing of the curvature tensor will guarantee the existence of the above mentioned vector field. From Eq.(5.4) and the symmetry of the Christoffel symbols, it follows that

$$A_{\mu, \beta} - A_{\beta, \mu} = \Gamma_{\beta\mu}^{\alpha} A_{\alpha}   - \Gamma_{\mu\beta}^{\alpha} A_{\alpha} = 0 \eqno(5.6) $$

Implying that $A_{\mu}$ is a gradient of a scalar $\phi$\ : $A_{\mu} = \phi_{,\mu}$. 

Thus if  $R^{\mu}_{\rho\alpha\beta}  = 0$, one can generate a vector field by parallel transport and the  generated field is the gradient of a scalar. Let' s choose new four scalars $\phi^{(\alpha)} $ and generate four parallel transported fields, $ A_{\mu} ^{(\alpha)}= \phi_{, \mu} ^{(\alpha)}$. Eq.(5.4) now yields

$$\phi_{,\mu,\nu}^{(\alpha)} = \Gamma_{\mu\nu}^{\beta} \phi_{,\beta}^{(\alpha)} \eqno(5.7) $$

Let's define new coordinates $\bar x^{\alpha} = \phi^{(\alpha)}$. Using the transformation law for the Christoffel symbols, Eq.(3.12), we find that 
$\bar\Gamma_{\mu\nu}^{\beta} =0$, namely $\bar g_{\mu\nu,\beta =0 }$
globally. Thus, the spacetime is flat and further linear transformation with constant coefficients, would yield coordinates $\{\widetilde x^{\alpha}\}$ in which $\widetilde g_{\mu\nu} = \eta_{\mu\nu}$, globally.

Since  $ R^{\mu}_{\rho\alpha\beta}  $ is a tensor, then from its transformation law it follows that if it vanishes  in {\it one} coordinate system, it  vanishes in {\it all} coordinates systems. Conversely, if it has at least one nonzero component in one system, it cannot vanish in any other coordinate system. In particular, even in a local inertial coordinate system in which $g_{\mu\nu} = \eta_{\mu\nu}$ and $g_{\mu\nu , \alpha} = 0$, it is impossible {\it even locally} to make  
e  $ R^{\mu}_{\rho\alpha\beta}  =0$. The reason is that the curvature tensor depends also on the second order derivatives of the metric that are nonzero even in a local inertial coordinate system. Therefore, the vanishing of the the curvature tensor is a coordinate invariant criterion for spacetime flatness.

\subsection*{Symmetries of the Curvature Tensor}

From eqs. (5.1) and (5.2) it follows that:

\paragraph{\hskip 1 truecm (i)} 

$$R_{\mu\nu\alpha\beta} = - R_{ \nu\mu\alpha\beta} = -R_{\mu\nu \beta\alpha} \eqno(5.8) $$
i.e. the curvature  tensor is antisymmetric in the two first and the two last indices.

\paragraph{\hskip 1 truecm (ii)} 

$$R_{\mu\nu\alpha\beta} =  R_{\alpha\beta\mu\nu}  \eqno(5.9)$$

i.e. the tensor is symmetric under the exchange of the first and last pairs of indices.
\paragraph{\hskip 1 truecm (iii)} 

$$R_{\mu\nu\alpha\beta}  +  R_{\mu \alpha\beta\nu} + R_{\mu  \beta\nu\alpha}  = 0 \eqno(5.10) $$

This last symmetry which is valid on its own, can be also derived from the two previous ones, except in the case where all indices have different values. In this case, the symmetry holds but is not a result of the previous ones.

$$R_{0123} + R_{0231} + R_{0312} = 0 \eqno(5.11)$$

The above symmetries leave us with 20 independent components of the curvature tensor (out of a total 64).  For further classification of the symmetries of the curvature tensor see the detailed discussion in Article 2.

\subsection*{Curvature Tensor of a Weak Static Gravitational Field}

Consider the metric corresponding to a weak static gravitational field as in section 2d. 

 $$ g_{\mu\nu} = \eta_{\mu\nu} + h_{\mu\nu} \ ;   \ \ h_{0i} = 0 \ ; \ \ h_{\mu\nu, 0} = 0 \ ;\ \  |h_{\mu\nu}| << 1  \eqno(5.12)$$

To first order in $h_{\mu\nu}$, the nonzero components of the curvature tensor are computed  to be,
$$R_{0i0j} = \frac{1}{2} h_{00,i,j} \eqno(5.13) $$
i.e. there are six independent nonzero components. With the identification of Eq.(2.19) ($h_{00} = \frac{2}{c^2}\phi $), Eq.(5.13)  becomes

$$R_{0i0j} = \frac{1}{c^2}\phi_{,i, j} \eqno( 5.14) $$

Consider two point masses separated by a spatial vector $d^j$. The tidal force per unit mass   exerted on the two masses, by a gravitational field $\phi$ is 

$$dF_i = F_{i, j} d^j = - \phi_{,i, j}d^j = - c^2 R_{0i0j} d^j \eqno(5.15) $$

Thus the curvature tensor represents gravitational tidal forces, at a given point. These forces cannot be eliminated by transforming to a local inertial coordinate system.

 \subsection*{The Bianchi Identities}
The curvature tensor can be shown to satisfy  a set  of four differential identities

$$R_{\mu\alpha\beta\gamma;\delta} +  R_{\mu\alpha \gamma\delta;\beta}             +R_{\mu\alpha\delta\beta;\gamma} = 0   \eqno(5.16) $$

This can be most easily proved by using Eq.(5.2) in a local inertial coordinate system.

 \subsection*{The Ricci Tensor and the Ricci Scalar}

Contracting the curvature tensor leads to the Ricci tensor $R_{\alpha\beta}$

$$ R_{\alpha\beta} \equiv R_{\alpha\mu\beta}^{\mu} \eqno(5.17)$$

The Ricci tensor is symmetric, due to the symmetries of the curvature tensor i.e. ,

$$R_{\alpha\beta} = R_{\beta\alpha} \eqno(5.18) $$

The Ricci scalar is obtained by further contraction of the Ricci tensor

$$ R\equiv R_{\alpha}^{\alpha} \eqno(5.19) $$

\subsection*{The Einstein Tensor and the  Einstein Scalar}

The Einstein tensor is defined by 
$$G_{\mu\nu} \equiv R_{\mu\nu} - \frac{1}{2}R g_{\mu\nu} \eqno(5.20)$$
and is evidently a symmetric tensor.

Contracting the Bianchi identities , Eq.(5.16)   with  $g^{\mu\beta} g^{\alpha\gamma}$
one finds

$$R_{\delta;\mu}^{\mu} -\frac{1}{2}R_{\delta} = G_{\delta;\mu}^{\mu} = 0  $$
and since the  covariant  derivative of the metric tensor vanishes identically, one gets

$$ G_{;\nu}^{\mu\nu} = 0  \eqno(5.21)$$

The Einstein scalar is obtained by contracting the Einstein tensor:
$$G\equiv G_{\alpha}^{\alpha} = - R \eqno (5.22) $$ 

\section*{6. Physics in a Curved Spacetime}

\subsection*{6a. General Principles}

How should physics be formulated in a curved spacetime? Given the form of  a physical equation in the SR framework, how should  one modify it in order to describe  properly the situation in the presence of gravity? The clue is provided by the general principles outlined in Chapter 2. The requirements of general covariance implies that physical quantities be represented by tensors and that the physical equations  be tensorial equations. 

Thus, let us introduce a local inertial coordinate system in which SR applies, and express the physical quantities, under consideration, and the equations satisfied by them
 in a four dimensional Minkowski tensorial form.  The equations can be considered as   the manifestation of the curved spacetime equation in the local inertial coordinates.

Since in the local inertial coordinates $g_{\mu\nu} = \eta_{\mu\nu}$ and covariant derivatives reduce to ordinary partial derivatives, the curved space form of the equation is readily obtained by replacing $\eta_{\mu\nu}$  by $g_{\mu\nu}$ and ordinary derivatives by the corresponding covariant derivatives. These guidelines are applied, in some detail, in the following sections.  

\subsection*{6b. Electromagnetic Field Equations}

The SR equations are

$$F_{,\nu}^{\mu\nu} = j^{\mu}\eqno(6.1) $$

$$F_{\mu\nu} = A_{\nu, \mu} -  A_{\mu, \nu}  \eqno(6.2)  $$

where $F^{\mu\nu}$ is the antisymmetric electromagnetic field tensor, derived from a vector potential $A^{\mu}$ and $j^{\mu} $  is the source current four vector. A charged particle of mass $m$ and charge $q$
moves under the influence of the field along trajectories given by

$$ u_{, \beta}^{\alpha} u^{\beta} = \frac{q}{m}F^{\alpha\gamma}u_{\gamma} \eqno(6.3)   $$ 

with $u^{\alpha}\equiv \frac{d x^{\alpha}}{ds}$ being the particle four velocity.

Applying the rules outlined in the previous section, we write the curved space equations as

$$F^{\mu\nu}_{;\nu}= j^{\mu} \eqno(6.4)$$

$$F_{\mu\nu} = A_{\nu; \mu} -  A_{\mu; \nu}  =  A_{\nu, \mu} -  A_{\mu, \nu}  \eqno(6.5)  $$

$$ u_{; \beta}^{\alpha} u^{\beta} = \frac{q}{m}F^{\alpha\gamma}u_{\gamma} \eqno(6.6)   $$ 

In Eq.(6.5) the antisymmetry of $F_{\mu\nu}$ was used. As expected, Eq.(6.6) reduces to a geodesic for neutral particles.

\subsection*{6c. Massive Scalar Field Equations}

The SR equation for a neutral massive scalar field $\phi$ of mass $m$ is

$$ \eta^{\alpha\beta} \phi_{,\alpha,\beta} + m^2 \phi = 0 \eqno(6.7)$$

which allows one to write the curved space equations as

$$g^{\alpha\beta} \phi_{;\alpha;\beta} + m^2 \phi = 0 \eqno(6.8)$$
Eq.(6.8) can possess solutions which are very different than those of Eq.(6.7) , due to the interaction with the gravitational field represented by $g_{\mu\nu}$. This interaction is of importance whenever the gravitational field   changes on   spatial or temporal scales which are comparable to or smaller than the characteristic a wavelength or period of  $\phi$. 

In particular, when $\phi$ represents a quantum field, this interaction can lead to particle creation. The latter will occur if gravity changes on spatial scales of the order of or smaller than the Compton length  or on timescales shorter than the Compton length divided by $c$.  
 Such situations exist near the horizons of black holes and in the early universe at times close to the cosmological singularity.

\subsection*{6d. Ideal Fluid equations}

An ideal fluid is defined as nonviscous, not sustaining shear dresses and not transferring heat between fluid elements. IN SR, the fluid is represented by an energy-stress tensor $T^{\mu\nu}$ which depends on the two scalars: the energy density $\rho$ and the pressure $p$, and on the flow four velocity vector
$u^{\alpha} = \frac{d x^{\alpha}}{ds}$,
 
$$ T^{\mu\nu} = (\rho + p) u^{\mu} u^{\nu} - p g^{\mu\nu} \eqno(6.9)$$

 $T^{\mu\nu}$ satisfies the energy-momentum conservation law
 
 $$T^{\mu\nu}_{,\nu} = 0 \eqno(6.10) $$
 
 The curved space form of  $T^{\mu\nu}$ is the same as that of Eq.(6.9) but the conservation law becomes
 
  $$T^{\mu\nu}_{;\nu} = 0 \eqno(6.11) $$

which upon substituting Eq.(6.9) leads to
$$ \rho_{,\nu} u^{\nu} + (\rho + p) u^{\nu}_{;\nu} = 0  \eqno(6.12) $$

and 

$$ u^{\mu}_{; \nu}  u^{\nu} + \frac{p_{,\nu}}{\rho + p} \left( u^{\mu} u^{\nu} - g^{\mu\nu}\right) = 0\eqno(6.13) $$ 

Eq.(6.12) expresses energy conservation while eqs. (6.13) are the equations of motion. Note that for dust particles, $p = 0$ Eq.(6.130 reduces to a geodesic. This is expected since $p = 0$ means that no interactions between the fluid particles are present, i.e. they are moving freely under the influence of gravity.

\subsection*{6e. Action Principle for Fields in a Curved spacetime}

Action principles are commonly used in the framework of SR to derive field equations. The action $I_m$ is a Minkowski invariant scalar

$$ I_m = \int L_m\left( \phi^{(r)} (x^{\alpha}),\phi^{(r)}_{, \mu}(x^{\alpha})\right) d^4x   \eqno(6.14) $$

where $L_m$ is the Lagrangian density which is a scalar depending on the fields  and their first derivatives. Upon arbitrary variations of the field $\phi^{(r)}$ that vanish on the boundary of the integration domain, one has

$$\delta I_m = \int \left(\frac{\partial L_m}{\partial\phi^{(r)} }  -   \left (\frac{\partial L_m}{\partial\phi^{(r)}_{,\alpha} }\right)_{,\alpha}   \right)\delta \phi^{(r)}d^4x \eqno(6.15)  $$  

By requiring that $\delta I_m = 0$, the field equations 

$$   \frac{\partial L_m}{\partial\phi^{(r)} }  -   \left (\frac{\partial L_m}{\partial\phi^{(r)}_{,\alpha} }\right)_{,\alpha}  =0 \eqno(6.16) $$
follow. 
As examples of Lagrangian densities, we give those for the electromagnetic field  and for the massive neutral scale field.

$$L_{EM} = \frac{1}{16 \pi} F_{\alpha\beta}F^{\alpha\beta}\ \ ; \ \ \   \ \ \    \ \ \   L_{scalar} =\frac{1}{8 \pi}  \phi_{,\alpha} \phi_{, \beta}\eta^{\alpha\beta} - m^2\phi^2$$
 
The Lagrangian density is used to define an energy-momentum tensor

$$T^{\mu\nu} = \eta^{\nu\alpha}  \frac{\partial L_m}{\partial\phi^{(r)}_{,\nu} }\phi^{(r)}_{,\alpha} -L_m \eta^{\mu\nu} \eqno(6.17)$$

Which satisfies the conservation law $T^{\mu\nu}_{, \nu} =0$. The energy-momentum tensor given by Eq.(6.17) is sometimes not symmetric and should be symmetrizing while taking care not to spoil $T^{\mu\nu}_{, \nu} =0$.
In curved spacetime the action is written as 

$$ I_m = \int\sqrt{-g} L_m   d^4x   \eqno(6.18) $$

where $L_m$ is a scalar ($\sqrt{-g} L_m$ is a scalar density which is obtained from the SR form of $L_m$ by replacing $\eta^{\mu\nu}$ with $g^{\mu\nu}$ 
and partial derivatives by covariant derivatives. As an example, consider the Lagrangian density for the massive scale neutral field

$$L_{scalar} =\frac{1}{8 \pi}( \phi_{;\alpha} \phi_{; \beta}\eta^{\alpha\beta} - m^2\phi^2)$$

The variation of the action due  to arbitrary variations in $\phi^{(r)}$ and in $g^{\mu\nu}$ is given by

$$\delta I_m = \int \sqrt{-g}\left(\frac{\partial L_m}{\partial\phi^{(r)} }  -   \left (\frac{\partial L_m}{\partial\phi^{(r)}_{,\alpha} }\right)_{,\alpha}   \right)\delta \phi^{(r)}d^4x + \frac{1}{2}\int\sqrt{-g} T_{\mu\nu}\delta g^{\mu\nu} d^4x \eqno(6.19)  $$

where $T_{\mu\nu}$ is the symmetric energy-momentum tensor

$$T_{\mu\nu} \equiv 2 \frac{\partial L_m}{\partial g^{\mu\nu}} -L_m g_{\mu\nu} \eqno(6.20) $$

Demanding that $\frac{\delta I_m}{\delta \phi^{(r)}} =0 $, one obtains the field equations
$$   \frac{\partial L_m}{\partial\phi^{(r)} }  - \frac{1}{\sqrt{-g}}  \left (\frac{\partial \sqrt{-g}L_m}{\partial\phi^{(r)}_{,\alpha} }\right)_{,\alpha}    =0 \eqno(6.21) $$

The variation of $I_m$ with respect to $g^{\mu\nu}$ is nonzero. Only the variation of the total action including also that of the gravitational field itself, does vanish with respect to variations of $g^{\mu\nu}$. We present the action for the gravitational field in Chapter 7. In the following section we show that the covariant conservation law $T^{\mu\nu}_{;\nu} = 0$ can be obtained as a result of $I_m$ being an invariant scalar.

\subsection*{6f. Conservation Laws in Curved spacetime}

\subsubsection*{Charge Conservation}

 From Eq.(6.4) one gets, using the symmetry of $F^{\mu\nu}$
$$j^{\mu}_{; \mu} = \frac{1}{\sqrt{-g}}\left( \sqrt{-g} j^{\mu} \right)_{,\mu} = \frac{1}{\sqrt{-g}}\left( \sqrt{-g}F^{\mu\nu}\right)_{,  \nu  , \mu}  =0 \eqno(6.22) $$
Thus giving rise  to a conserved scalar charge

$$Q =\int_{\Sigma} \sqrt{-g} j^{\alpha} d^3\Sigma_{\alpha} \eqno(6.23) $$

where $\Sigma$ is any space-like 3-surface which  at its spatial infinity $j^{\alpha}$ vanishes. When $\Sigma$ is normal to $x^0$, Eq.(6.23) becomes

$$Q =\int \sqrt{-g} j^0 d^3 x^{\alpha} \eqno(6.24) $$

Therefore, electric charge is conserved in the presence of gravity. In general, any conservation law of  a four vector $A^{\mu}$, which in SR reads $A^{\mu}_{,\mu} =0$ becomes in curves spacetime, $A^{\mu}_{;\mu} =0$,
leading to a conserved scalar  $\int\sqrt{-g} A^0 d^3x$. As we shall see this is not the case with conserved tensors.

 \subsubsection*{Energy-Momentum Tensor Conservation}

The action  $I_m$ is an invariant constant scalar and as such, does not change under coordinate transformations. Consider therefore an infinitesimal coordinate transformation of the type

$$ \bar x^{\mu} = x^{\mu} + \xi^{\mu} \ \ ; \ \ \ \ \ |\xi^{\mu}| << \ \ |x^{\mu}|  \eqno(6.25)$$

under which $\delta I_m = 0$ automatically. In Eq.(6.19), one has to insert $\delta g^{\mu\nu}$ and $\delta\phi^{(r)}$ corresponding to the transformation given by Eq.(6.25). The first integral in Eq.(6.19) 
will vanish automatically  as a result of the field equations, eq.(6.21). The second integral, though nonzero  in general, {\it does } vanish for a $\delta g^{\mu\nu}$ induced by a coordinate transformation due to $I_m$ being an invariant scalar. 

The variation of $g^{\mu\nu}$ resulting from the coordinate transformation of Eq.(6.25) is computed to be

$$\delta g^{\mu\nu} =\bar g^{\mu\nu}(x^{\alpha }) - g^{\mu\nu}(x^{\alpha })=
\xi^{\mu ; \nu} +  \xi^{\nu ; \mu} \eqno(6.26) $$

where $\xi^{\mu ; \nu} \equiv   \xi^{\mu}_{ ;\alpha} g^{\alpha \nu}$. As the second integral of eq.(6.21) must vanish for $\delta g^{\mu\nu}$ given by Eq.(6.26), we obtain, making use of the symmetry of $T^{\mu\nu}$

$$T^{\mu \nu} _{ ; \nu} = 0 \eqno(6.28) $$ 

This result shows that the conservation law of the energy-momentum tensor is a direct consequence of the invariance of the action to coordinate transformations.

It is well known that in the framework of SR, the conservation law of the energy-momentum tensor leads to a conserved four vector of energy-momentum for isolated systems. $T^{\mu\nu}_{, \nu} = 0$ implies that

$$\frac{\partial}{\partial x^0}
\int T^{\mu 0} d^3 x = - \int T^{\mu i}_{, i}  d^3x \eqno(6.29) $$

Since the right hand side is a divergence, it can be expressed as an integral over a 2-surface enclosing the system. For an isolated system $T^{\mu\nu} = 0$ at spatial  infinity  so that the 2-surface integral vanishes and one has a conserved Minkowski four vector of energy-momentum

$$P^{\mu} = \int T^{\mu 0} d^3 x  \eqno(6.30) $$

In the case of  curved spacetime, the situation is quite different. Eq.(6.28) can be written as 

$$\left(\sqrt{-g}T^{\mu\nu}\
\right)_{, \nu} + \sqrt{-g} T^{\alpha\beta} \Gamma^{\mu}_{\alpha\beta} = 0 \eqno(6.31) $$

Therefore,  $\int \sqrt{-g}  T^{\mu 0} d^3 x$ is not a conserved quantity. This signals that the energy-momentum of the gravitational field itself must be taken into account if one wishes  to have a conserved energy-momentum  four vector for an isolated system. This topic is discussed in Chapter 8.

\section*{7. Einstein Field Equations}
\subsection*{7a.  Derivation by Correspondence with Newtonian Gravity in the Limit of a Static Weak Gravitational Field}

In the previous chapters we dealt with the effects of gravity on matter and non gravitational physics. We address now the issue of the field equations for gravity itself. In the framework of the geometric representation of gravity, we are interested in finding the field equations that determine  the geometry of spacetime.  A guideline towards the formulation of the equations is provided by considering the case of a static weak field  generated by a non relativistic ideal fluid, i.e. one for which  $p<< \rho$. Upon using eqs. (5.13), ( 5.14) , one finds that 

$$R_{00} = -\frac{1}{2}\nabla^2h_{00} = -\frac{1}{c^2}\nabla^2 \phi     \eqno(7.1) $$

where $\phi$ is the Newtonian gravitational potential that obeys the Poisson equation

$$\nabla^2 \phi = \frac{4\pi G}{c^2}\rho =  \frac{4\pi G}{c^2}T_{00}\eqno(7.2) $$
In Eq.(7.2) use has been made of the fact that $p<< \rho$. Combining eqs. (7.1) an Eq.(7.2) yields

$$R_{00} =  -\frac{4\pi G}{c^4}T_{00} \eqno(7.3) $$

Let's regard Eq.(7.3) as the static weak field limit 
of the sought for tensorial equations. Eq.(7.3)  suggests that the source term be given by the energy-momentum tensor $T_{\mu\nu}$ of matter and material fields.  It also suggests that the geometric tensor on the left hand side be  a second order symmetric tensor  made out of combinations of the curvature tensor. This ensures linearity in the second derivatives of  $g_{\mu\nu}$. Since $T^{\mu\nu}$  is conserved, the geometric tensor must be one that is conserved by the virtue of its structure. Furthermore, demanding that SR be a possible empty-space solution, we require that in the case of $T_{\mu\nu} = 0$ , the flat spacetime solution will be a possible one. These requirements single out the sought for tensor as the Einstein tensor, $G_{\mu\nu}$, introduced in Chapter 5. Thus, the Einstein field equations are

$$G_{\mu\nu} = - \frac{8\pi G}{c^4}T_{\mu\nu} \eqno(7.4) $$

where the proportionality constant was determined by the demand that Eq.(7.3) be the weak field limit of Eq.(7.4).

The field equations may be generalized by adding a "cosmological constant"
term $\Lambda g_{\mu\nu} $ to the left hand side so that
$$G_{\mu\nu} + \Lambda g_{\mu\nu}= - \frac{8\pi G}{c^4}T_{\mu\nu} \eqno(7.5) $$

The generalized field equations Eq.(7.5) are compatible with $T^{\mu\nu}_{; \nu} = 0$ but do not  admit a flat spacetime as a possible matter-vacuum ( $T^{\mu\nu} = 0$) solution. For $T^{\mu\nu} =0$ Eq.(7.5) implies $R =4\Lambda$ thus at least one component of the curvature tensor is nonzero. By transferring this term to right hand side one can interpret $\frac{c^4}{8\pi G}\Lambda g_{\mu\nu}$ as the energy-momentum of the vacuum, the so called {\it Dark Energy} favored in recent years by cosmological observations that indicate an accelerated expansion of the universe. The observational upper bounds on $\Lambda$ are $|\Lambda|  \les  10^{-56} cm^{-2}$. In what follows we adopt $\Lambda = 0$.

The field equations Eq.(7.4) imply that $T^{\mu\nu}$ is automatically conserved. This follows since $G^{\mu\nu}_{;\nu} = 0$
independently, as a result of the Bianchi identities, Eq.(5.16). Thus, the conservation of the energy-momentum tensor becomes a consequence  of the way matter is coupled to gravity. As $T^{\mu\nu}_{; \nu} = 0$ contain the the equations of motion  of matter, it follows that the equations of motion of the matter sources of gravity are contained in the field equations. This situation is rather different from that in electrodynamics where the  field equations contain only the charge conservation law, Eq.(6.22)  but not the equations of motion, Eq.(6.6) which are independent of the field equations, Eqs.(6.4), (6.5). As a matter of fact, in electrodynamics one could 
alter arbitrary the equations of motion, while taking care  to satisfy charge conservation, regardless of the electromagnetic field equations.

  In gravity, this is no longer possible: the field equations dictate very specific equations of motion. This feature  of the Einstein field equations manifests itself in analyzing the initial value problem, concerning the data specified on a spacelike three surface, $x^0 =  const$, that determine the future time evolution of the gravitational field and matter and matter fields. An analysis shows that one cannot prescribe arbitrary initial values for  the matter variables, the $g_{\mu\nu}$, and their first order time derivatives, but there are four constraints that the initial data must satisfy. A detailed discussion of these aspects is given in Article 13.
  
  The field equations, Eq.(7.4), are ten in number. However, as a result of the four Bianchi identities there are only six independent equations. Therefore, the solution of the ten components   of $g_{\mu\nu}$ contains four unspecified functions. This enables assigning arbitrary values to four of the ten  $g_{\mu\nu}$ components. In fact, this situation is expected and reflects the freedom to  perform arbitrary coordinate transformations. Such transformations do not change the nature of the  geometry and that of the gravitational field represented by this geometry.  They do however    introduce four unspecified functiones  defining the new coordinate system in terms of the old ones. Only if one specifies a given coordinate system, this indeterminacy will be removed.
 
 In particular, useful coordinate conditions are those defining "harmonic coordinates". Define the coordinates to be given by four scalars, so that $x^{\mu} = \phi^{(\mu)}$, that satisfy the curved spacetime harmonic equation (hence the name)
 
 $$\Box \phi^{(\mu)} = \frac{1}{\sqrt{- g}}\left(\sqrt{- g} \phi^{(\mu)}_{, \alpha}g^{\alpha\lambda}\right)_{, \lambda} = 0  \eqno(7.6)     $$
 
In terms of the coordinates $x^{\mu} = \phi^{(\mu)} $ Eq.(7.16) takes the form

$$\frac{1}{\sqrt{- g}}\left(\sqrt{- g}   g^{\mu\lambda}\right)_{, \lambda} =  - g^{\lambda\tau} \Gamma^{\mu}_{\lambda\tau} = 0  \eqno(7.7)  $$

Eq.(7.7) defines the harmonic coordinates. In order to actually construct the harmonic coordinates, let us use the transformation law for the Christoffel symbols, Eq.(3.12),  to derive the transformation law for $ g^{\lambda\tau} \Gamma^{\mu}_{\lambda\tau}$, i.e.    
 
$$ \bar g^{\lambda\tau}(\bar x^{\alpha})\bar\Gamma^{\mu}_{\lambda\tau}(\bar x^{\alpha}) = g^{\tau\sigma} \Gamma^{\rho}_{\tau\sigma} \frac{\partial \bar x^{\mu}}{\partial x^{\rho}} - g^{\rho\sigma} \frac{\partial}{\partial x^{\sigma}} \left(\frac{\partial \bar x^{\mu}}{\partial x^{\rho}}\right)\eqno(7.8) $$

 Starting with non harmonic coordinates $(x^{\alpha})$, we can ensure that 
  $(\bar x^{\alpha})$ be harmonic by demanding the right hand-side of Eq.(7.8) to vanish. This in turn  yields four differential equations for $ \bar x^{\mu}( x^{\alpha})$  that could be solved provided  that $det( g^{\mu\nu}) \neq 0$. This is assumed to hold anyway, since it is the condition that $g_{\mu\nu}$ be defined once  $g^{\mu\nu} $ are given. We make use of the harmonic coordinates in chapter 9 when dealing in the weak field limit of the field equations.
  
 \subsection*{7b. Action Principle for the Gravitational Field}  
  
 It would be desirable to have an action principle from which the field equations Eq.(7.4) could be derived. The action for the matter sources $I_m$ was treated in 
 Chapter 6. We are now looking for an action for the gravitational field, $I_g$
 
 $$I_g = \int \sqrt{-g} L_g d^4 x \eqno(7.9) $$ 
  
  where $L_g$ is  a scalar, so that

  $$\delta I_G = \frac{c^4}{16 \pi G}\int \sqrt{-g} G_{\mu\mu}\delta g^{\mu\nu} d^4 x = 0     \eqno(7.10)   $$
    
  This will ensure that the total action $I = I_g + I_m$ will have  a variation  
  
$$  \frac{1}{2}\int\sqrt{-g}\left( \frac{c^4}{16 \pi G}   G_{\mu\mu}  + T_{\mu\nu}\right)\delta g^{\mu\nu} d^4x + \int \sqrt{-g}\left(\frac{\partial L_m}{\partial\phi^{(r)} }  -   \left (\frac{\partial L_m}{\partial\phi^{(r)}_{,\alpha} }\right)_{,\alpha}   \right)\delta \phi^{(r)}d^4x    \eqno(7.11) $$

for variations $\delta g^{\mu\nu}\ {\rm  and\ }$ $\delta \phi^{(r)}$ that vanish on  the three surface bounding the integration domain of Eq.(7.11). Upon requiring that $\delta I$ vanishes with respect  to these variations, we obtain the matter equations, Eq.(6.21) and the Einstein field equations, Eq.(7.4). Let us turn now to the Lagrangian density, $L_g$. One cannot form a scalar out of $g^{\mu\nu}$ and their first derivatives  other than a constant number. A scalar can however be formed that contains also the second derivatives of the metric and is {\it linear} in them.  The most general such scalar is a linear combination of the Ricci tensor $R = R^{\alpha}_{\alpha}$ and a constant number. A short computation yields

$$\delta\int \sqrt{-g} R d^4x =  \int \sqrt{-g} G_{\mu\nu} \delta g_{\mu\nu} d^4 x + \int (\sqrt{-g} v^{\alpha})_{, \alpha} d^4 x \eqno(7.12)   $$

with 

$$v^{\alpha} = g^{\rho\alpha} \delta \left( \Gamma^{\beta}_{\beta\rho}\right) - g^{\rho\beta} \delta\left(\Gamma ^{\alpha}_{\beta\rho}\right)  $$ 

The second integral on the right hand-side of Eq.(7.12) can be expressed a   3-surface integral of $\sqrt{-g} v^{\alpha}$ over the boundary of the integration domain of Eq.(7.12). On this boundary, the variation in
$g^{\mu\nu}$ is zero and therefore $v^{\alpha}$ vanishes there. Thus, by comparing Eq.(7.10) with Eq.(7.12) the sought for $L_g$ is found to be

$$L_g = \frac{c^4}{16 \pi G}R \eqno(7.13) $$

Note that since $I_g$ is an invariant scalar, its value does not change under coordinate transformations. Therefore, in analogy with the analysis given in Section 6f, we find that  $G^{\mu\nu}_{; \nu} = 0$. This relation does not contain a new information, however, since  $G^{\mu\nu}_{; \nu} = 0$ identically as result of the Bianchi  identities.

We note that the field equations containing the cosmological term, Eq.(7.5) can be derived if $I_g$ is modified to 

$$I_g^* = \frac{c^4}{16 \pi G}\int\sqrt{-g} ( R - 2 \Lambda) d^4 x  $$

The action $I_g$ can be split into two parts, one containing only $g_{\mu\nu}$ and their first derivatives and the other containing also second  derivatives of $g_{\mu\nu}$. The splitting is based on separating $\sqrt{-g}R$ as follows 

$$\frac{c^4}{16 \pi G}R = A^{\alpha}_{,\alpha} + \sqrt{-g} \widetilde L_g \eqno(7.14) $$

where

 $$\frac{16 \pi G}{c^4} A^{\alpha} = \sqrt{-g} ( g^{\rho\alpha} \Gamma^{\beta}_{\beta\rho} - g^{\rho\tau} \Gamma^{\beta}_{\rho\tau}) \eqno(7.15) $$ 

and
 
 $$\frac{16 \pi G}{c^4}\widetilde L_g =g^{\rho\beta} (\Gamma^{\alpha }_{\tau\alpha} (\Gamma^{\tau}_{\rho\beta} -
  \Gamma^{\tau}_{\alpha\rho} \Gamma^{\alpha}_{\tau\beta}) \eqno(7.16) $$
  
  Clearly, $\widetilde L_g$ contains only $g_{\mu\nu}$ and their first derivatives. The action $L_g$ is therefore expressible as
  
 $$ I_g = \int\sqrt{-g} \widetilde L_g d^4 x + \int A^{\alpha} d^3 S_{\alpha} \eqno(7.17) $$
  
  Upon variation of $g_{\mu\nu}$ that vanish on the 3-surface bounding the integration domain, the second integral on the right hand-side of Eq.(7.17) has a zero variation. Thus,
  
$$\delta I_g = \delta \int \sqrt{-g} \widetilde L_g d^4 x = \frac{c^4}{16 \pi G}\int\sqrt{-g} G_{\mu\nu} \delta g^{\mu\nu} d^4 x  \eqno(7.18) $$
  
$$ \frac{c^4}{16 \pi G} G_{\mu\nu}  = \frac{\partial(\sqrt{-g} \widetilde L_g)}{\partial g^{\mu\nu}} 
  - \left(\frac{\partial(\sqrt{-g} \widetilde L_g)}{\partial g^{\mu\nu}_{ , \alpha}}\right)_{, \alpha} \eqno(7.19) $$
  
  Note that $\widetilde L_g$ is not a true scalar and it can be made to vanish by transforming to a local inertial coordinate system. However, $\delta(\sqrt{-g} \widetilde L_g)$ { \bf  is}  a true scalar density. $\widetilde L_g$ is useful in constructing the gravitational energy-momentum pseudo tensor discussed in Chapter 8.

\section*{8. Gravitational Energy-Momentum Pseudo-Tensor}

In Chapter 6 we found that the covariant conservation law of the matter energy-momentum tensor, $T^{\mu\nu}$, does not lead to a conserved four vector of energy-momentum   for an isolated system. This indicates that the energy-momentum content  of the gravitational field itself should also be accounted for in the total energy-momentum balance. Thus, one is motivated to construct a quantity $t^{\mu}_{\nu}$ such that

$$ \left(\sqrt{-g} (T^{\nu}_{\mu} +t^{\nu}_{\mu} )\right)_{, \nu} = 0 \eqno(8.1) $$

where  $t^{\mu}_{\nu}$  is formed from $g_{\mu\nu}$ and their derivatives. In order to find   $t^{\mu}_{\nu}$, let's express the covariant conservations law $T^{\mu\nu}_{; \nu} = 0$ in the form

$$\frac{1}{\sqrt{-g}}(\sqrt{-g} T^{\nu}_{\mu})_{, \nu} - \frac{1 }{2} g_{\alpha\beta , \mu}T^{\alpha\beta} = 0 \eqno(8.2)$$

 which upon using the field equation, Eq.(7.4) can be rewritten as 

$$\frac{1}{\sqrt{-g}}(\sqrt{-g} T^{\nu}_{\mu})_{, \nu} - \frac{c^4 }{16\pi G} G_{\alpha\beta} g^{\alpha\beta}_{, \mu} = 0 \eqno(8.3)$$ 

Using Eq.(7.19), we find that Eq.(8.1) can be satisfied if $t^{\nu}_{\mu}$ is given by

$$t^{\nu}_{\mu} = g^{\alpha\beta}_{, \mu} \frac{\partial \widetilde L_g}{\partial g^{\alpha\beta}_{, \nu}}
- \delta^{\nu}_{\mu} \widetilde L_g \eqno(8.4) $$

Where $\widetilde L_g$ is given by Eq.(7.16). Note that $t^{\nu}_{\mu}$ is formed out of $g_{\mu\nu}$ and their first derivatives and is quadratic in the first derivatives.

Even though one may be tempted to interpret  $t^{\nu}_{\mu}$ as the energy-momentum density of the gravitational field, this is not the case. Unlike $T^{\nu}_{\mu}$, $t^{\nu}_{\mu}$ is not a genuine tensor since it is formed out of non tensors. Thus, even in the presence of gravity, one can make $t^{\nu}_{\mu} = 0$
by transforming to a local inertial coordinate system. On the other hand in a flat spacetime where there is no gravity present, one can obtain $t^{\nu}_{\mu} \neq 0$  by transforming to non Lorentzian coordinates. These, imply that there is no meaning to local gravitational energy density. This in full accord with the EP that states that any gravitational field can be transformed a way locally, by transforming to a local inertial coordinate system.

Nonetheless, one can define meaningful quantity representing the total energy-momentum of an isolated system, provided spacetime is asymptotically flat at spatial infinity. To do so, note that Eq.(8.1) implies that

$$\frac{\partial}{\partial x^0} \int \sqrt{-g} (T^0_{\mu} + t^0_{\mu} ) d^3 x = \int \sqrt{-g} (T^i_{\mu} + t^i_{\mu} ) d^2 S_i = 0 \eqno(8.5)$$
where the 2-surface integral is over a boundary enclosing the subsystem, at spatial infinity. For a spacetime that is flat at spatial infinity, and where no radiation escapes through the boundary, then in Euclidean coordinates  (in which $g_{\mu\nu} = const$), $T^{\nu}_{\mu} = t^{\nu}_{\mu}= 0 $ on the boundary and the surface integral vanishes, giving rise to a conserved quantity

$$P_{\mu} = \int \sqrt{-g} (T^0_{\mu} + t^0_{\mu} ) d^3 x \eqno(8.6) $$ 

$P_{\mu}$ is invariant under arbitrary coordinate transformations that do not change the Euclidean coordinates at spatial infinity. Under arbitrary  coordinate transformations that   at spatial infinity tend to Lorentz Transformations, $P_{\mu}$transforms like  Minkowski four vector. Thus, one may regard 
$P_{\mu}$ as the total energy-momentum  four vector of the isolated system. Note however, that the existence of 
$P_{\mu}$ critically depends on spacetime being flat at spatial infinity.

\section*{9. Weak Gravitational Field -- the Linearized Field Equations}

In many situations, the gravitational field is weak and the metric can be written as

$$ g_{\mu\nu} = \eta_{\mu\nu}+h_{\mu\nu} \ \ \ \ ; \ \ \ \ \ \ \ \  |h_{\mu\nu}|<< 1\eqno(9.1)   $$

In this case it is useful to apply the linear approximation to the Einstein field equations, Eq.(7.4), in which only terms up to first order in $h_{\mu\nu}$ are retained. In order to obtain the linearized field equations, Let us write down the Riemann curvature tensor Eq.(5.2) which in this approximation is given by

$$R_{\mu\rho\alpha\beta} = \frac{1}{2}( h_{\alpha\mu, \rho, \beta} - h_{\beta\mu, \rho, \alpha} + h_{\beta\rho, \mu , \alpha} - h_{\alpha\rho, \mu, \beta}) \eqno(9.2) $$

From which, the computed Ricci tensor follows

$$R_{\rho\beta} = \eta^{\mu\alpha}R_{\mu\rho\alpha\beta} =   \frac{1}{2}( h_{ , \rho, \beta} - h^{\alpha}_{\beta , \rho, \alpha} + h_{\beta\rho, \mu , \alpha } \eta^{\mu\alpha}  -  h^{\alpha}_{\rho, \alpha , \beta})\eqno(9.3) $$  

where

$$h \equiv \eta^{\mu\nu} h_{\mu\nu} \eqno (9.4) $$

Note that since $g_{\mu\alpha} g^{\alpha\nu} = \delta^{\nu}_{\mu}$ Eq.(9.1) implies that

$$g^{\mu\nu} = \eta^{\mu\nu} - h^{\mu\nu} \eqno(9.5) $$

where

$$h^{\mu\nu} = \eta^{\mu\alpha} \eta^{\nu\beta} h_{\alpha\beta} \eqno(9.6)  $$

A substantial simplification in the expressions of $R_{\mu\nu}$ and $G_{\mu\nu}$ is achieved when harmonic coordinates, Eq.(7.7), are employed. For  a weak field, Eq.(7.7) reduces to

$$h^{{\hskip -0.1 cm }^*\mu\nu}_{, \nu} = 0   \eqno(9.7) $$

where 

 $$ h^{{\hskip -0.1 cm }^*\mu\nu} \equiv h_{\mu\nu} -  \frac{1}{2}\eta_{\mu\nu} h  \eqno(9.8) $$
 
 With Eq.(9.7) satisfied, we obtain  using Eq.(9.3), that 
 
 $$ R_{\mu\nu} = \frac{1}{2} \eta^{\alpha\beta} h_{\mu\nu, \alpha, \beta} = \frac{1}{2}  \Box
  h_{\mu\nu} \eqno(9.9) $$

and 
$$G_{\mu\nu} = \frac{1}{2}  \Box
 h^{{\hskip -0.1 cm }^* }_{\mu\nu}  \eqno(9.10) $$

Therefore, the linearized field equations are 

$$ \Box
 h^{{\hskip -0.1 cm }^* }_{\mu\nu} = -\frac{16 \pi G}{c^4}T_{\mu\nu} \eqno(9.11)   $$

The automatic conservation law $ G^{\mu\nu}_{; \nu} = 0$ reduces in the weak field case to

$$ G^{\mu\nu}_{; \nu} = \Box h^{{\hskip -0.1 cm }^*\mu\nu}_{, \nu} = 0 \eqno(9.12) $$

which is indeed compatible with the harmonic  condition Eq.(9.7). The energy-momentum conservation law reduces to the flat space form $T^{\mu\nu}_{, \nu} = 0$. Thus, in the linear approximation the gravitational field potentials  $ h^{{\hskip -0.1 cm }^*\mu\nu}$ obey an inhomogeneous wave equation. This is consistent with neglecting effects due to the energy-momentum density of the gravitational field itself. Such a neglect is justified in this case since $t_{\mu\nu}$, Eq.(8.4) is of second order in $h_{\mu\nu}$.

Does the harmonic coordinate condition, Eq.(9.7), fix uniquely the coordinate system? To answer this question, let's consider an infinitesimal coordinate transformation {\hskip 2truecm }  $x^{\mu} \rightarrow \bar x^{\mu} = x^{\mu} + \xi^{\mu}$ where
$| \xi^{\mu}|<< |x^{\mu}|$. To first order in $ \xi^{\mu}$, we find

$$\bar g_{\mu\nu} = \eta_{\mu\nu}+ h_{\mu\nu} - \xi_{\mu, \nu} - \xi_{\nu, \mu} \equiv \eta_{\mu\nu} + \bar h_{\mu\nu} \eqno(9.13) $$  

where

$$\bar h_{\mu\nu} \equiv h_{\mu\nu} - (\xi_{\mu, \nu} + \xi_{\nu, \mu})\ \ \ \  ; \ \ \ \ \ \ \ \xi_{\mu} \equiv \eta_{\mu\nu} \xi^{\nu} \eqno(9.14) $$

From Eq.(9.14) 

$$ \bar {h^{{\hskip -0.1 cm }^* }}_{\mu\nu} = h_{\mu\nu}  - (\xi_{\mu, \nu} + \xi_{\nu, \mu}) + \eta_{\mu\nu} \xi^{\alpha}_{, \alpha} \eqno(9.15) $$

so that

$$ \bar {h^{{\hskip -0.1 cm }^* }}^{\mu\nu}_{, \nu} =  h^{{\hskip -0.1 cm }^*\mu\nu}_{, \nu} - \Box \xi^{\mu} \
 eqno(9.16) $$

Therefore, if the original coordinates were harmonic, so will be thenew coordinates, provided that $\Box \xi^{\mu} =0$. Thus, the harmonic coordinates are fixed up to $\xi^{\mu}$ that satisfy the homogeneous wave equation.

The weak field approximation presented here is applied in Article 5 in discussing gravitational  waves.

  \section*{Acknowledgement} 

 Thanks are due to Idit Goldman for an expert proof reading of the manuscript.

\end{document}